# On the Computation of the Higher-Order Statistics of the Channel Capacity for Amplify-and-Forward Multihop Transmission

Ferkan Yilmaz, *Member, IEEE*, Hina Tabassum, *Student, IEEE*,

and Mohamed-Slim Alouini, *Fellow, IEEE*

**Abstract**

Higher-order statistics (HOS) of the channel capacity provide useful information regarding the level of reliability of the signal transmission at a particular rate. We propose in this letter a novel and unified analysis, which is based on the moment-generating function (MGF) approach, to efficiently and accurately compute the HOS of the channel capacity for amplify-and-forward multihop transmission over generalized fading channels. More precisely, our mathematical formulism is easy-to-use and tractable specifically requiring only the reciprocal MGFs of the instantaneous signal-to-noise ratio distributions of the transmission hops. Numerical and simulation results, performed to exemplify the usefulness of the proposed MGF-based analysis, are shown to be in perfect agreement.

**Index Terms**

Higher-order statistics, ergodic capacity, multihop transmission, amplify-and-forward transmission, generalized fading channels.

## I. INTRODUCTION

Relay technology has emerged as one of the the key enablers, according to the latest and most recent technologies and standardizations [1], to achieve not only *higher quality* but also *higher reliability* in

Manuscript received August 13, 2012. The review of this paper was coordinated by Xxxxxxxx.

Ferkan Yilmaz, Hina Tabassum and Mohamed-Slim Alouini are with King Abdullah University of Science and Technology (KAUST), Al-Khawarizmi Applied Math. Building, Thuwal 23955-6900, Makkah Province, Saudi Arabia (e-mail: {ferkan.yilmaz, hina.tabassum, slim.alouini}@kaust.edu.sa).

This work was supported by KAUST.

Digital Object Identifier 10.1109/XXX.2012.xxxxxx.xxxxxx

throughput as well as larger coverage and lower latency in wireless signal transmission since the relay technology is an effective method to combat pathloss and fading [2]. Accordingly for wireless signal transmission, amplify-and-forward (AF) multihop transmission has incessantly attracted a great deal of attention of the researchers in the last decade [1], [2].

The ergodic capacity is one of the information-theoretic performance metrics to quantify the achievable maximum throughput in wireless signal transmission. Within this context, some recent analysis frameworks [3]–[5] have appeared in the literature to quantify the ergodic capacity of AF multihop transmission over generalized fading channels. In particular, Farhadi and Beaulieu elegantly proposed in [3] a characteristic function (CF)-based approach. Later, Waqar *et al.* proposed in [5] an approximation-based approach using deformed-Stieltjes transform applied on the truncated series expansion of the logarithm function. Finally, Yilmaz *et al.* introduced in [4] a moment-generating function (MGF)-based approach to efficiently and accurately compute the ergodic capacity of AF multihop transmission over generalized fading channels. Nonetheless, these analysis frameworks are not enough to fully explore *the reliability of the signal transmission*. More precisely, and referring to the fact that the reliability is the ability of an item to perform a required function under stated conditions for a stated period of time, the reliability of AF multihop transmission is mainly characterized by the degree of dispersion or scatter in the channel capacity, rather than by the average of the channel capacity.

In a reliable transmission, the maximum dispersion or the peak-and-trough in the channel capacity should be controlled in order to achieve a successful carrier aggregation [1]. It is hence very useful to obtain the higher-order statistics (HOS) of the channel capacity. Several papers have addressed the HOS of the channel capacity for different types of fading channels [6]–[8, and references there in]. In particular, while some of the references in [6] have dealt with multiple input multiple output (MIMO) transmission over Rayleigh or Rician fading channels, Laourine *et al.* considered in [6] the HOS of the channel capacity over log-normal fading channels. There has been indeed a lack of unified analysis in the literature on the HOS of the channel capacity. Within this fact, Di Renzo *et al.* have presented in [7] a numerical approach to evaluate the HOS of the channel capacity for maximal ratio combining (MRC) diversity receivers. However, as mentioned in [7, Section III-E], this numerical solution requires both higher-order differentiation and a limit operation. In order to circumvent these computational difficulties, [8] offered recently an MGF-based approach for the HOS of the channel capacity. While the approaches developed in [7], [8] are applicable to MRC diversity receivers, they are neither applicable nor extendible to the computation of the HOS of the channel capacity for the AF multihop transmission, and to the best of our knowledge, there is no unified analysis available in the literature investigating this subject. In this letter, we propose a unified MGF-based approach to efficiently and accurately compute the HOS of the channel capacity for AF multihop transmission over generalized fading channels.

## II. HIGHER-ORDER STATISTICS OF THE CHANNEL CAPACITY FOR AF MULTIHOP TRANSMISSION

The end-to-end instantaneous signal-to-noise ratio (SNR) $\gamma_{end}$ of AF multihop transmission over generalized fading channels is simply characterized by the normalized harmonic mean of the transmission hops' instantaneous SNRs, that is given by [2, Eq. (4)]

$$\gamma_{end} = \frac{1}{\frac{1}{\gamma_1} + \frac{1}{\gamma_2} + \ldots + \frac{1}{\gamma_L}}, \tag{1}$$

where $L$ is the number of hops in the multihop transmission and $\gamma_\ell$ denotes the instantaneous SNR of the $\ell$th hop. Note that the reciprocal distribution of $\gamma_{end}$ is defined by $\widetilde{\gamma}_{end} = 1/\gamma_{end}$ and can be readily re-written as $\widetilde{\gamma}_{end} = \sum_{\ell=1}^{L} \widetilde{\gamma}_\ell$ where $\widetilde{\gamma}_\ell$ is the reciprocal distribution of the $\ell$th hop's instantaneous SNR $\gamma_\ell$. Hence, the higher-order statistics of the channel capacity for AF multihop transmission, i.e., $\mu_n = \mathbb{E}[\log^n(1+\gamma_{end})]$, where $\mathbb{E}[\cdot]$ denotes the expectation operator, $n \in \mathbb{N}$ is the order of the statistics and $\log(\cdot)$ represents the natural logarithm, can be obviously re-written as $\mu_n = \mathbb{E}[\log^n(1+1/\widetilde{\gamma}_{end})] = \int_0^\infty \log^n(1+1/\gamma) \, p_{\widetilde{\gamma}_{end}}(\gamma) \, d\gamma$, where $p_{\widetilde{\gamma}_{end}}(\gamma)$ is the probability density function (PDF) of $\widetilde{\gamma}_{end}$ and typically not available in closed-form when $L$ is greater than two. In general, $\mu_n$ is given by

$$\mu_n = \underbrace{\int_0^\infty \int_0^\infty \ldots \int_0^\infty}_{L\text{-fold}} \log^n\left(1 + \left(\sum_{\ell=1}^{L} \gamma_\ell\right)^{-1}\right) \, p_{\widetilde{\gamma}_1, \widetilde{\gamma}_2, \ldots, \widetilde{\gamma}_L}(\gamma_1, \gamma_2, \ldots, \gamma_L) \, d\gamma_1 d\gamma_2 \ldots d\gamma_L, \tag{2}$$

where $p_{\widetilde{\gamma}_1, \widetilde{\gamma}_2, \ldots, \widetilde{\gamma}_L}(\gamma_1, \gamma_2, \ldots, \gamma_L)$ is the joint PDF of the reciprocal distributions $\widetilde{\gamma}_1, \widetilde{\gamma}_2, \ldots, \widetilde{\gamma}_L$. It should be mentioned that (2) is computationally cumbersome, especially as the number of hops $L$ is greater than two. Further, it cannot be separated into a product of one dimensional integrals because of the non-linearity of the natural logarithm $\log(\cdot)$. In order to circumvent these difficulties, a novel MGF-based approach is introduced in the following theorem requiring only the reciprocal MGFs of the hops.

TABLE I
RECIPROCAL MGFs OF SOME WELL-KNOWN CHANNEL FADING MODELS AND THEIR FIRST-ORDER DERIVATIVES

| **Instantaneous SNR Model, i.e., $p_{\gamma_\ell}(\gamma)$** | **Reciprocal MGF $\mathcal{M}_{\widetilde{\gamma}_\ell}(s)$ and its First-Order Derivative $\frac{\partial}{\partial s}\mathcal{M}_{\widetilde{\gamma}_\ell}(s)$** |
|---|---|
| **Gamma Fading Environment** [2] $p_{\gamma_\ell}(\gamma) = \frac{1}{\Gamma(m_\ell)}\left(\frac{m_\ell}{\bar{\gamma}_\ell}\right)^{m_\ell}\gamma^{m_\ell-1}\exp\left(-\frac{m_\ell}{\bar{\gamma}_\ell}\gamma\right),$ | $\mathcal{M}_{\widetilde{\gamma}_\ell}(s) = \frac{2}{\Gamma(m_\ell)}\left(\frac{m_\ell s}{\bar{\gamma}_\ell}\right)^{\frac{m_\ell}{2}} K_{m_\ell}\left(2\sqrt{\frac{m_\ell s}{\bar{\gamma}_\ell}}\right),$ $\frac{\partial}{\partial s}\mathcal{M}_{\widetilde{\gamma}_\ell}(s) = -\frac{2}{\Gamma(m_\ell)}\left(\frac{m_\ell s}{\bar{\gamma}_\ell}\right)^{\frac{m_\ell+1}{2}} s^{\frac{m_\ell}{2}} K_{m_\ell-1}\left(2\sqrt{\frac{m_\ell s}{\bar{\gamma}_\ell}}\right),$ |

where the fading parameters $m_\ell \geq 1/2$ and $\bar{\gamma}_\ell > 0$ are the fading figure and average power, respectively. Moreover, $K_m(\cdot)$ is the $m$th-order modified Bessel function of second kind [9, Eq.(18.16)].

| **Generalized Gamma Fading Environment** [10] $p_{\gamma_\ell}(\gamma) = \frac{\xi_\ell}{\Gamma(m_\ell)}\left(\frac{\beta_\ell}{\bar{\gamma}_\ell}\right)^{m_\ell\xi_\ell}\gamma^{m_\ell\xi_\ell-1}\exp\left(-(\beta_\ell/\bar{\gamma}_\ell)^{\xi_\ell}\gamma^{\xi_\ell}\right)$ | $\mathcal{M}_{\widetilde{\gamma}_\ell}(s) = \frac{1}{\Gamma(m_\ell)}\Gamma\left(m_\ell, 0, \frac{\beta_\ell}{\bar{\gamma}_\ell}s, \frac{1}{\xi_\ell}\right),$ $\frac{\partial}{\partial s}\mathcal{M}_{\widetilde{\gamma}_\ell}(s) = -\frac{\beta_\ell}{\Gamma(m_\ell)\,\bar{\gamma}_\ell}\Gamma\left(m_\ell - \frac{1}{\xi_\ell}, 0, \frac{\beta_\ell}{\bar{\gamma}_\ell}s, \frac{1}{\xi_\ell}\right),$ |

where the fading parameters $m_\ell$ ($0.5 \leq m_\ell < \infty$), $\xi_\ell$ ($0 \leq \xi_\ell < \infty$) and $\bar{\gamma}_\ell > 0$ are the fading figure, shaping factor and average power, respectively. Moreover, $\beta_\ell = \Gamma(m_\ell + 1/\xi_\ell)/\Gamma(m_\ell)$, and $\Gamma(\cdot, \cdot, \cdot, \cdot)$ is the extended incomplete Gamma function [11, Eq.(6.2)]. Note that the special or limiting cases of GNM fading are well-known in the literature including Rayleigh, half-normal (i.e., half-Gaussian), Nakagami-$m$, Weibull, lognormal, and AWGN.

## A. General Result on the HOS of the Channel Capacity

**Theorem 1.** *The HOS of the channel capacity for AF multihop transmission over generalized fading channels is given by*

$$\mu_n = \int_0^\infty \mathcal{Z}_n(s)\left\{\mathcal{M}_{\widetilde{\gamma}_{end}}(s) - \frac{\partial}{\partial s}\mathcal{M}_{\widetilde{\gamma}_{end}}(s)\right\}ds, \tag{3}$$

*where $n \in \mathbb{N}$ denotes the order of the statistics, and where $\mathcal{M}_{\widetilde{\gamma}_{end}}(s) = \mathbb{E}[\exp(-s\,\widetilde{\gamma}_{end})]$ is the reciprocal MGF of the end-to-end instantaneous SNR $\gamma_{end}$. Furthermore, the auxiliary function $\mathcal{Z}_n(s)$ is given by*[1,2]

$$\mathcal{Z}_n(s) = (-1)^n {}_1F_1^{(n,0,0)}[1;1;-s], \tag{4}$$

*where for $m, n, q \in \mathbb{N}$ and $a, b, x \in \mathbb{R}$, ${}_1F_1^{(m,n,q)}[a;b;x]$ denotes Kummer's differentiated confluent hypergeometric (DCH) function defined in [12, Eqs. (07.20.20.0010.02), (07.20.20.0011.02), and (07.20.20.0012.02)].*

*Proof:* Note that using the relation between the differentiation and the $n$th power of logarithm [9, Eq. (5.1.4/(i))], we can readily write

$$\frac{\partial^n}{\partial a^n}\frac{\widetilde{\gamma}_{end}^a}{(1+\widetilde{\gamma}_{end})^a} = \frac{(-1)^n\,\widetilde{\gamma}_{end}^a}{(1+\widetilde{\gamma}_{end})^a}\log^n(1+\gamma_{end}). \tag{5}$$

---
[1] Note that the auxiliary function $\mathcal{Z}_n(s)$ is a closed-form expression and is implemented as a built-in function in MATHEMATICA®, that is
$$Z[n\_, s\_] := (-1)^n \text{ Hypergeometric1F1}^{(n,0,0)}[1, 1, -s];$$
It can be moreover efficiently computed using numerical differentiation in MAPLE® and MATLAB™ mathematical packages.

Then, $\mu_n = \mathbb{E}[\log^n(1+\gamma_{end})]$ can be represented as

$$\mu_n = (-1)^n \frac{\partial^n}{\partial a^n} \int_0^\infty \frac{\gamma^a}{(1+\gamma)^a} p_{\widetilde{\gamma}_{end}}(\gamma)\, d\gamma \bigg|_{a=0}. \tag{6}$$

By means of using the integral equality [13, Eq. (3.35.1/3)], the integral representation of $\gamma^a/(1+\gamma)^{a+1}$ can be given by

$$\frac{\gamma^a}{(1+\gamma)^{a+1}} = \int_0^\infty \exp(-\gamma u) \,_1F_1[1+a;1;-u]\, du. \tag{7}$$

Then, substituting (7) into (6) and utilizing the Kummer's DCH function [12, Eq. (07.20.20.0010.02)] results in

$$\mu_n = (-1)^n \int_0^\infty {}_1F_1^{(n,0,0)}[1;1;-u]\left\{\int_0^\infty (1+\gamma)\exp(-\gamma u)\, p_{\widetilde{\gamma}_{end}}(\gamma)\, d\gamma\right\} du. \tag{8}$$

Substituting the definition of the reciprocal MGF, i.e., $\mathcal{M}_{\widetilde{\gamma}_{end}}(u) = \int_0^\infty \exp(-\gamma u)\, p_{\widetilde{\gamma}_{end}}(\gamma)\, d\gamma$ into the inner integral easily results in (11), which proves Theorem 1. ∎

Under the well-justified assumption that there does not exist any correlation among all reciprocal instantaneous SNRs $\widetilde{\gamma}_1, \widetilde{\gamma}_2, \ldots, \widetilde{\gamma}_L$, the HOS of the channel capacity can be easily computed by using the following corollary.

**Corollary 1.** *When there is no correlation among all hops, the HOS of the channel capacity is given by*[2]

$$\mu_n = \int_0^\infty \mathcal{Z}_n(s)\left\{1 - \sum_{\ell=1}^L \frac{(\partial/\partial s)\mathcal{M}_{\widetilde{\gamma}_\ell}(s)}{\mathcal{M}_{\widetilde{\gamma}_\ell}(s)}\right\}\prod_{\ell=1}^L \mathcal{M}_{\widetilde{\gamma}_\ell}(s)\, ds, \tag{9}$$

*where $\mathcal{M}_{\widetilde{\gamma}_\ell}(s)$ and $(\partial/\partial s)\mathcal{M}_{\widetilde{\gamma}_\ell}(s)$ denote the reciprocal MGF of the $\ell$th hop's instantaneous SNR and its first-order derivative, respectively, and are available in closed-form for a variety of fading channel models as exemplified in Table I.*

*Proof:* When there is no correlation among hops, one can readily write the joint MGF $\mathcal{M}_{\widetilde{\gamma}_{end}}(s) = \prod_{\ell=1}^L \mathcal{M}_{\widetilde{\gamma}_\ell}(s)$. Then the proof is obvious using the derivative of $\mathcal{M}_{\widetilde{\gamma}_{end}}(s)$. ∎

---

[2] It is worth emphasizing that the main goal in a MGF-based approach is to convert the multidimensional integration of interest into a product of single-dimensional integrations each of which can be obtained in a closed form using Laplace transform [14]. More specifically, after using an MGF-based approach, the obtained analytical result should have one single-integration, and the obtained analytical result should be then efficiently computed using only one single-quadrature rule. With these observations, the developed approach in Theorem 1 and Corollary 1 is evidently an MGF-based approach.

*B. Special Cases of the Auxiliary Function $\mathcal{Z}_n(s)$ and Some Statistical Applications of the HOS of the Channel Capacity*

Note that substituting $n = 1$ into (4) and then performing some algebraic manipulations, one can reduce $\mathcal{Z}_1(s)$ to

$$\mathcal{Z}_1(s) = -e^{-s}\big(\mathbf{E} - \mathrm{Ei}(s) + \log(s)\big), \quad (10)$$

where the constant $\mathbf{E} = 0.5772156649015328606...$ is the Euler-Mascheroni constant [12] and $\mathrm{Ei}(\cdot)$ is the exponential integral function [12, Eq. (06.35.07.0001.01)]. The first-order statistics of the channel capacity, i.e., the ergodic capacity $\mu_1 = \mathbb{E}[\log(1 + \gamma_{end})]$ is then efficiently computed by

$$\mu_1 = -\int_0^\infty e^{-s}\big(\mathbf{E} - \mathrm{Ei}(s) + \log(s)\big)\left\{\mathcal{M}_{\widetilde{\gamma}_{end}}(s) - \frac{\partial}{\partial s}\mathcal{M}_{\widetilde{\gamma}_{end}}(s)\right\}ds, \quad (11)$$

which is numerically equivalent to [4, Eq. (9)] but converging faster; within this respect, it may be useful for the researchers working in this field.

The variance of the channel capacity is a measure of how far the channel capacity lies from the ergodic capacity and related to the first and second-order statistics of the channel capacity by $\mathrm{var} = \mu_2 - \mu_1^2$. Furthermore, its normalization with respect to the ergodic capacity corresponds to the amount of dispersion (AoD) in the channel capacity, specifically given by $\mathrm{AoD} = \mu_2/\mu_1 - \mu_1$ taking values between zero and one[3]. Within this context, the reliability of the throughput can be defined as $\mathrm{R} = 100 - 100\,\mathrm{AoD}$. Accordingly, substituting $n = 2$ into (4) results in

$$\mathcal{Z}_2(s) = e^{-s}\bigg(\mathbf{E}^2 - \frac{\pi^2}{6} - 2\,\mathbf{E}\,\mathrm{Ei}(s) + 2\big(\mathbf{E} + \mathrm{Ei}(s)\big)\log(s) - 3\log^2(s)\bigg) + \sum_{n=0}^{2}\binom{2}{n}\log^n(s)\,\mathcal{Q}_\infty(s; 2-n), \quad (12)$$

where $\mathcal{Q}_n(x;\nu)$, $x \in \mathbb{C}$ and $n,\nu \in \mathbb{N}$ is a computationally efficient and recurrent polynomial defined by

$$\mathcal{Q}_n(x;\nu) = \mathcal{Q}_{n-1}(s;\nu) + x^{(n-1)/2}\,Q_{n-1}(\nu), \quad (13\mathrm{a})$$

$$Q_n(\nu) = \frac{4}{n^2}\frac{\Gamma^{(\nu)}(1+n/2)}{\Gamma^{(\nu)}(n/2)}Q_{n-2}(\nu), \quad (13\mathrm{b})$$

with the initial conditions $\mathcal{Q}_0(x;\nu) = 0$, $Q_0(\nu) = \Gamma^{(\nu)}(1)$ and $Q_1(\nu) = 0$, where $\Gamma^{(\nu)}(\cdot)$ is the $\nu$-order derivative of the Gamma function [12]. Next, substituting (12) into (11), the second-order statistics of the channel capacity is efficiently computed. Accordingly, the variance and the AoD in the channel capacity can also be readily obtained.

---
[3]Note that for generalized one-sided Gaussian, Weibull, generalized Gamma and extended generalized-K fading environments, there is no physical justification, to the best of our knowledge, in the literature that the shape parameter $\xi$ (i.e., not fading figure/ diversity order $m$) can be smaller than one.

The other two important statistical metrics of the channel capacity are the skewness $\text{S} = (\mu_3 - \mu_1^3)/(\mu_2 - \mu_1^2)^{3/2}$ and the kurtosis $\text{K} = (\mu_4 - \mu_1^4)/(\mu_2 - \mu_1^2)^2$. More specifically, the skewness corresponds to a measure of the degree of asymmetry of the distribution of the channel capacity while the kurtosis to the degree of peakedness of the channel capacity around the ergodic capacity. It is apparent that in order to obtain the skewness and kurtosis, the third- and fourth-order statistics are strictly required; and for this purpose, (4) can be easily and efficiently computed. Within this context, it is worth mentioning that the higher-order statistics can also be efficiently and accurately computed by means of Theorem 1 utilized with the following theorem.

**Theorem 2.** *A computationally efficient, numerically accurate and extremely tight approximation of the auxiliary function $\mathcal{Z}_n(s)$ is given by*

$$\mathcal{Z}_n(s) = \sum_{k=0}^{n} \frac{(-1)^k}{2\Delta^n} \binom{n}{k} \left\{ L_{-1-k\Delta}(-s) + (-1)^n L_{-1+k\Delta}(-s) \right\}, \quad (14)$$

*where $L_n(\cdot)$ is the Laguerre polynomial of order $n$ [12, Eq. (07.02.02.0001.01)] and $\Delta$ is a small number (e.g., $\Delta \approx 0.01$).*

*Proof:* Proof is obvious using the Grunwald-Letnikov differentiation [15] and the Laguerre polynomial representation of the hypergeometric function (i.e., $_1F_1[1+a;1;z] = L_{-1-a}(z)$ where $a, z \in \mathbb{R}$) [12, Eq. (07.20.03.0009.01)]. ∎

### III. NUMERICAL AND SIMULATION RESULTS

For multihop transmission in generalized Gamma fading environments, the HOS of the channel capacity can be easily and efficiently computed using Theorem 1 as well as Corollary 1, yielding the equation given at the top of the next page. In order to check its analytical simplicity and accuracy with the exact auxiliary function in (4) and its tight approximation in (14), the HOS of the channel capacity for the AF triple-hop transmission is depicted in Fig. 1, in which numerical and simulation results are shown to be in perfect agreement. Accordingly, the variance, skewness and kurtosis of the AF multihop transmission can be easily evaluated using (15). As further seen in Fig. 1, the HOS of the channel capacity are almost the same around $7\,\text{dB}$. For the average SNR lower than $7\,\text{dB}$, the multihop transmission significantly gets

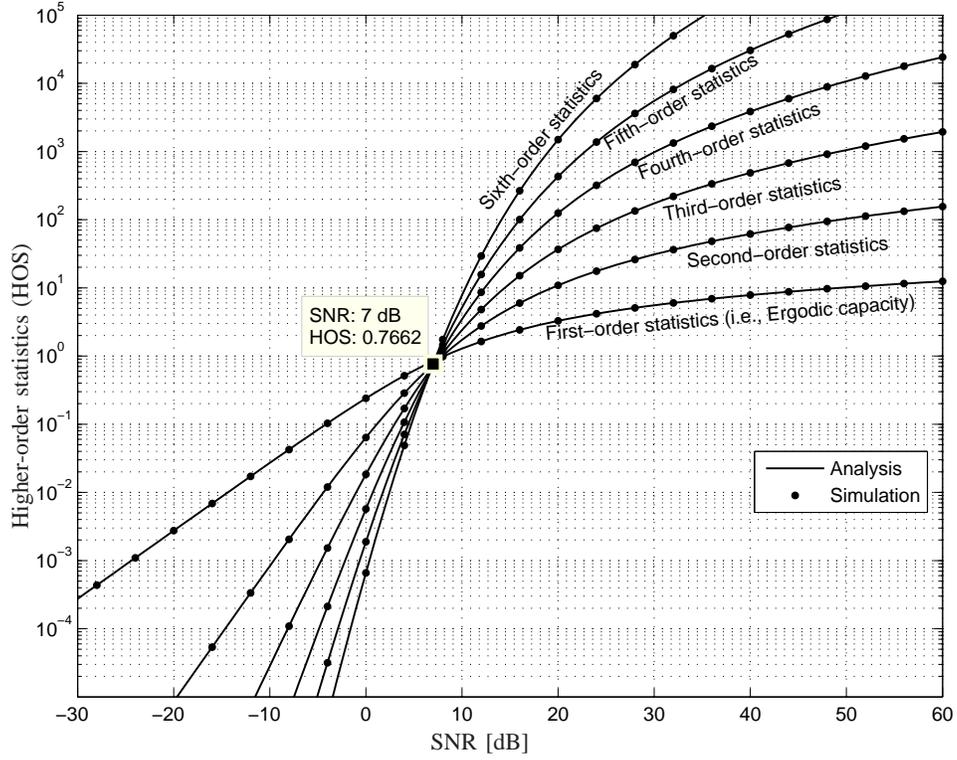

Fig. 1. HOS of the channel capacity for AF triple-hop transmission ($L = 3$) over generalized Gamma fading channels. The shape parameter and fading figure are chosen the same as $\xi_\ell = 1.23$ and $m_\ell = 2.34$ for all $\ell \in \{1, 2, 3\}$.

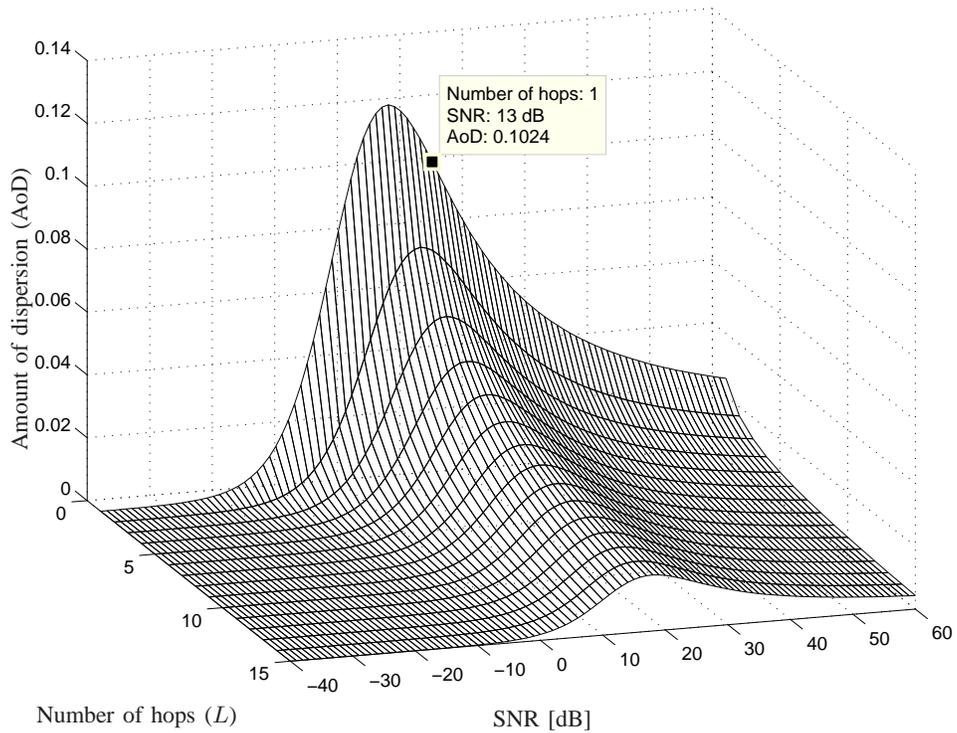

Fig. 2. AoD in the channel capacity for AF multihop transmission over generalized Gamma fading channels. The shape parameter and fading figure are chosen the same as $\xi_\ell = 1.23$ and $m_\ell = 2.34$ for all $\ell \in \{1, 2, \ldots, L\}$.

$$\mu_n = \int_0^\infty \mathcal{Z}_n(s) \left\{ \prod_{\ell=1}^L \frac{1}{\Gamma(m_\ell)} \Gamma\left(m_\ell, 0, \frac{\beta_\ell}{\bar{\gamma}_\ell} s, \frac{1}{\xi_\ell}\right) \right\} \sum_{\ell=1}^L \frac{\beta_\ell}{\Gamma(m_\ell) \bar{\gamma}_\ell} \Gamma\left(m_\ell - \frac{1}{\xi_\ell}, 0, \frac{\beta_\ell}{\bar{\gamma}_\ell} s, \frac{1}{\xi_\ell}\right) \prod_{\substack{k=1 \\ k \neq \ell}}^L \frac{1}{\Gamma(m_\ell)} \Gamma\left(m_\ell, 0, \frac{\beta_\ell}{\bar{\gamma}_\ell} s, \frac{1}{\xi_\ell}\right) ds, \tag{15}$$

into the low-SNR regime, and the HOS of the channel capacity can be characterized in this regime with $\mu_n \approx \mu_1^n$. In particular, the variance of the channel capacity significantly goes to zero in low-SNR regime.

In addition, Fig. 2 depicts the behavior of the AoD in the channel capacity with respect to different number of hops and average SNR values. More precisely, as the average SNR increases, the AoD in the channel capacity for a specific number of hops sharply increases, reaches its peak value, and then slightly decreases. For low- and high-SNR values, the AoD distinctly diminishes, which means that the channel capacity does not change drastically and the throughput is reliable (i.e., stable around its mean value). Furthermore, the AoD decreases as the number of hops increases. The other important benefit of AF multihop transmission is therewith to prominently enhance the reliability of transmission. For a better communication configuration, the SNR should be consequently chosen greater than the SNR value for which the AoD peaks. As for example seen in Fig. 2, the AoD is $0.1024$ for $13$ dB in a single-hop transmission, and accordingly its reliability percentage is $R = 89.76$. In other words, either the average SNR should be chosen equal to or greater than $13$ dB or the number of hops should chosen equal to or greater than $2$ hops in order to reach at least $R = 100 - 100\,\text{AoD} = 89.76$ percent reliable transmission throughput.

## IV. Conclusion

In this letter, a novel MGF-based approach is proposed for the higher-order statistics of the channel capacity for AF multihop transmission over generalized fading channels. Our approach leads to a generic new expression which is computationally efficient and accurate and which requires only the reciprocal MGFs of the transmission hops. In addition, numerical results are shown to be in perfect agreement with those obtained via Monte-Carlo simulations, thus verifying the accuracy of our theoretical analysis.